\documentclass[fleqn,twoside]{article}
\usepackage{espcrc2}
\input{epsfig.sty}
\def\Journal#1#2#3#4{{#1} {\bf #2}, #3 (#4)}

\def\NPA{{\rm Nucl. Phys.} A}
\def\NPB{{\rm Nucl. Phys.} B}
\def\PLB{{\rm Phys. Lett.}  B}
\def\PRL{\rm Phys. Rev. Lett.}
\def\PRC{{\rm Phys. Rev.} C}
\def\PRD{{\rm Phys. Rev.} D}

\def\la{\langle}
\def\ra{\rangle}
\def\be{\begin{equation}}
\def\ee{\end{equation}}
\def\bea{\begin{eqnarray}}
\def\eea{\end{eqnarray}}
\def\lsim{\mathrel{\rlap{\lower4pt\hbox{\hskip1pt$\sim$}}
    \raise1pt\hbox{$<$}}}         
\def\gsim{\mathrel{\rlap{\lower4pt\hbox{\hskip1pt$\sim$}}
    \raise1pt\hbox{$>$}}}

\usepackage{graphicx}
\usepackage[figuresright]{rotating}

\newcommand{\AmS}{{\protect\the\textfont2
  A\kern-.1667em\lower.5ex\hbox{M}\kern-.125emS}}

\title{Pion Form Factor and Quark Mass Evolution in a
Light-Front Quark Model}

\author{Ho-Meoyng Choi\address[MCSD]{Department of Physics,
        Carnegie Mellon University, \\
        Pittsburgh, PA 15213, U.S.A}%
        \thanks{homeoyng@andrew.cmu.edu},
        L. S. Kisslinger\addressmark\thanks{
               kissling@andrew.cmu.edu}
        and
        Chueng-Ryong Ji\address{Department of Physics, North Carolina State
                         University,\\
        Raleigh, NC 27695-8202, U.S.A}
        \thanks{crji@unity.ncsu.edu}}
\begin{document}

\begin{abstract}
We discuss the soft contribution to the elastic pion form factor with
the mass evolution from current to constituent quark being taken into
account in a light-front quark model(LFQM).
\vspace{1pc}
\end{abstract}

\maketitle


 The pion electromagnetic (EM) form factor is of great interest for the
study of Quantum Chromodynamics (QCD). At low momentum transfers ($Q^2$)
nonperturbative QCD (NPQCD) dominates, while at large $Q^2$ perturbative
QCD (PQCD) can be used to calculate the asymptotic form factor; and the
transition from NPQCD to PQCD has long been of
interest.
The light-front (LF) quantization method~\cite{BPP} may be most useful
in connecting the formulations of NPQCD and PQCD since the LF
wavefunctions provide the essential link between hadronic phenomena at
short distances(perturbative) and at long distances(nonperturbative).
Although the relevant minimum momentum scale for the PQCD exclusive
processes is still under a debate~\cite{BJPR}, the LF method has been
successfully applied to the constituent quark model and described the
hadron properties at low momentum transfer region quite
well~\cite{JK,CJ}.
In many previous quark models~\cite{JK,CJ},
a constant constituent quark mass was used in the analysis of the hadron
properties especially at $Q^2 < 1$ GeV$^2$.
As shown in the literatures~\cite{JK,CJ}, such constituent quark
model has been quite successful in
describing static properties of a hadron such as the form factor, charge
radius, and decay constant etc..
On the other hand, the approach based on the quantum field theory
such as the Dyson Schwinger Equations(DSEs)~\cite{DS} uses the running
mass instead of constant constituent mass
and it also gives properties of the pion that are in agreement with the
experimental data.

Thus, in this talk, we present the quark mass
evolution effect on the pion in a light-front quark
model(LFQM)~\cite{KCJ}.
In the present work we restrict ourselves to the soft NPQCD
part with a LFQM, but an essential ingredient is the use of
a running quark mass, which is the main subject of this talk.

The form factor of the pion is related to the matrix element
of the current by the following equation:
\be\label{Current}
\la J^\mu_{e.m.}\ra\equiv\la P'|{\bar q}\Gamma^\mu q|P\ra
= (P' + P)^\mu F_\pi(Q^2).
\ee
In usual LF frame, the form factor of a hadron can be obtained by
the sum of valence and nonvalence diagrams.
However, if we choose the Drell-Yan-West(DYW)(or $q^+=0$) frame with
``$+$"-component of the current, only the valence diagram is needed. Then,
the matrix element of the current
given by Eq.~(\ref{Current}) can be expressed as a convolution integral
in terms of LF wave function, $\Psi(x,{\bf k}_\perp)$ as follows:
\bea\label{jmu}
\la J^\mu_{e.m}\ra &=&\sum_{\lambda_q\lambda'_q\lambda_{\bar q}}
\int^1_0 dx\int d^2{\bf k}_\perp
\Psi_{\lambda'_q\lambda{\bar q}}(x,{\bf k'}_\perp)\nonumber\\
&\times&\frac{{\bar u}_{\lambda'_q}(p'_q)}{\sqrt{p'^+_q}}\Gamma^\mu
\frac{u_{\lambda_q}(p_q)}{\sqrt{p^+_q}}
\Psi_{\lambda_q\lambda{\bar q}}(x,{\bf k}_\perp),
\eea
where $p^+_q$=$p'^+_q$=$(1-x)P^+$ and
${\bf k'}_\perp$=${\bf k}_\perp -x{\bf q}_\perp$ in the initial pion
rest frame, ${\bf P}_\perp$=0. The helicity of the quark(antiquark) is
denoted as $\lambda_{q({\bar q})}$.
In our model calculation of the pion form factor,
we use the Gaussian radial wave function as well as the relativistic
spin-orbit wave function obtained by the interaction independent Melosh
transformation(see~\cite{KCJ} for more detail).

In the usual light-front constituent quark model~\cite{JK,CJ},
the bare quark-photon vertex, $\Gamma^\mu=\gamma^\mu$, is used.
However, when the quark propagator has momentum-dependent dressing,
the bare vertex is no longer adequate because it
violates the Ward-Takahashi identity(WTI).
In general, the solution of the DSE for the renormalized dressed-quark
propagator takes the form $S(p)^{-1}=A(p^2){\not\! p} - B(p^2)$
in Minkowski space,
where the quark mass evolution function $m(p^2)$ is defined as
$m(p^2)=B(p^2)/A(p^2)$. Also, gauge invariance requires that the quark-photon
vertex $\Gamma^\mu$ given by Eq.~(\ref{jmu}) satisfy the WTI, i.e.
$-q^\mu\Gamma_\mu(p;q)= S(p')^{-1}-S(p)^{-1}$
( current conservation) as well as
$\Gamma^\mu(p;0)=\partial S(p)^{-1}/\partial p_\mu$
(charge conservation),
where $q=p - p'$.

As used in many DSE studies of EM interactions~\cite{DS},
we take the Ball-Chiu(BC) ansatz~\cite{BC} for
the quark-photon vertex
\bea\label{BC}
&&\hspace{-0.7cm}\Gamma^\mu_{\rm BC}=
\frac{ ({\not\! p}
+ {\not\! p'})}{2}(p + p')^\mu
\frac{ A(p'^2)-A(p^2)}{p'^2-p^2} \nonumber\\
&&\hspace{-0.7cm}+ \frac{A(p'^2)+A(p^2)}{2}\gamma^\mu
- (p + p')^\mu\frac{B(p'^2)-B(p^2)}{p'^2-p^2}.\nonumber\\
\eea
Although the asymptotic behavior of the running mass might require
crossing symmetry(under $Q^2\leftrightarrow-Q^2$) at high momentum
transfer, there is no clue yet for the low momentum transfer region.
So, we introduce two algebraic parametrizations of the running mass;
one satisfying the crossing symmetry(CS) and the
other satisfying the crossing asymmetry(CA):
\bea\label{CSCA}
m^{\rm CS}(p^4)=m_0 + (m_c - m_0)\frac{ 1+e^{-\mu^4/\lambda^4}}
{1 +e^{(p^4 - \mu^4)/\lambda^4}},&&
\nonumber\\
m^{\rm CA}(p^2)=m_0 + (m_c - m_0)\frac{ 1+e^{-\mu^2/\lambda^2}}
{1 +e^{(p^2 - \mu^2)/\lambda^2}},&&
\end{eqnarray}
where $m_0$ and $m_c$ are the current and constituent quark masses,
respectively. The parameters $\mu$ and $\lambda$ are used to adjust
the shape of the mass evolution.

\begin{figure}
\centerline{\psfig{figure=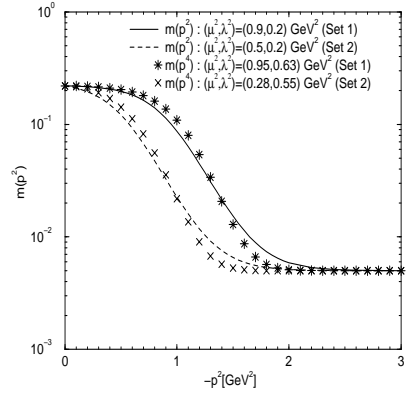,height=60mm,width=60mm}}
\vspace{-1cm}
\caption{Quark mass evolution in spacelike momentum region,
$-p^2>0$.\label{massfig}}
\end{figure}

For comparison, we use in Fig.~\ref{massfig} two different parameter
sets for each mass evolution function.
The current and constituent quark
masses used are $m_0=5$ MeV and $m_c=220$ MeV, respectively.
Simulating the constituent picture at the small momentum region,
we have chosen these particular sets of parameters, [Set 1] and [Set 2]
for each mass function, to keep the constituent mass up
to $(-p^2)\sim 1$ and 0.5 GeV$^2$, respectively, before it drops
exponentially.

In order to express the four momentum $p^2$ in terms of LF variables
$(x,{\bf k}_\perp)$, we use the on-mass shell condition, $p^2=m^2(p^2)$.
It implies zero binding energy of a mock meson, i.e.
$P^-=p^-_q + p^-_{\bar q}$ where $P^-(=P^0-P^3)$ and $p^-_q(p^-_{\bar q})$
are the LF energies of the mock meson and the quark(antiquark),
respectively. It leads to the following identity for the pion case
($m_q=m_{\bar q}$),  $p^2 = x(1-x){\tilde M}^2 - {\bf k}^2_\perp$.
For the mock meson mass ${\tilde M}$, we
take the average value(so called spin-averaged meson mass) of
$\pi(m_\pi)$ and $\rho(m_\rho)$ masses with appropriate weighting factors
from the spin degrees of freedom, i.e. ${\tilde M}$=$(m_\pi
+ 3m_\rho)_{\rm exp}/4$=612 MeV.

In our numerical calculations, we use the model parameters
$(m_c,\beta)$=(0.22,0.3659) [GeV] obtained in Ref.~\cite{CJ} for the
linear confining potential model.

\begin{figure}[t]
\vspace{-1cm}
\centerline{\epsfig{figure=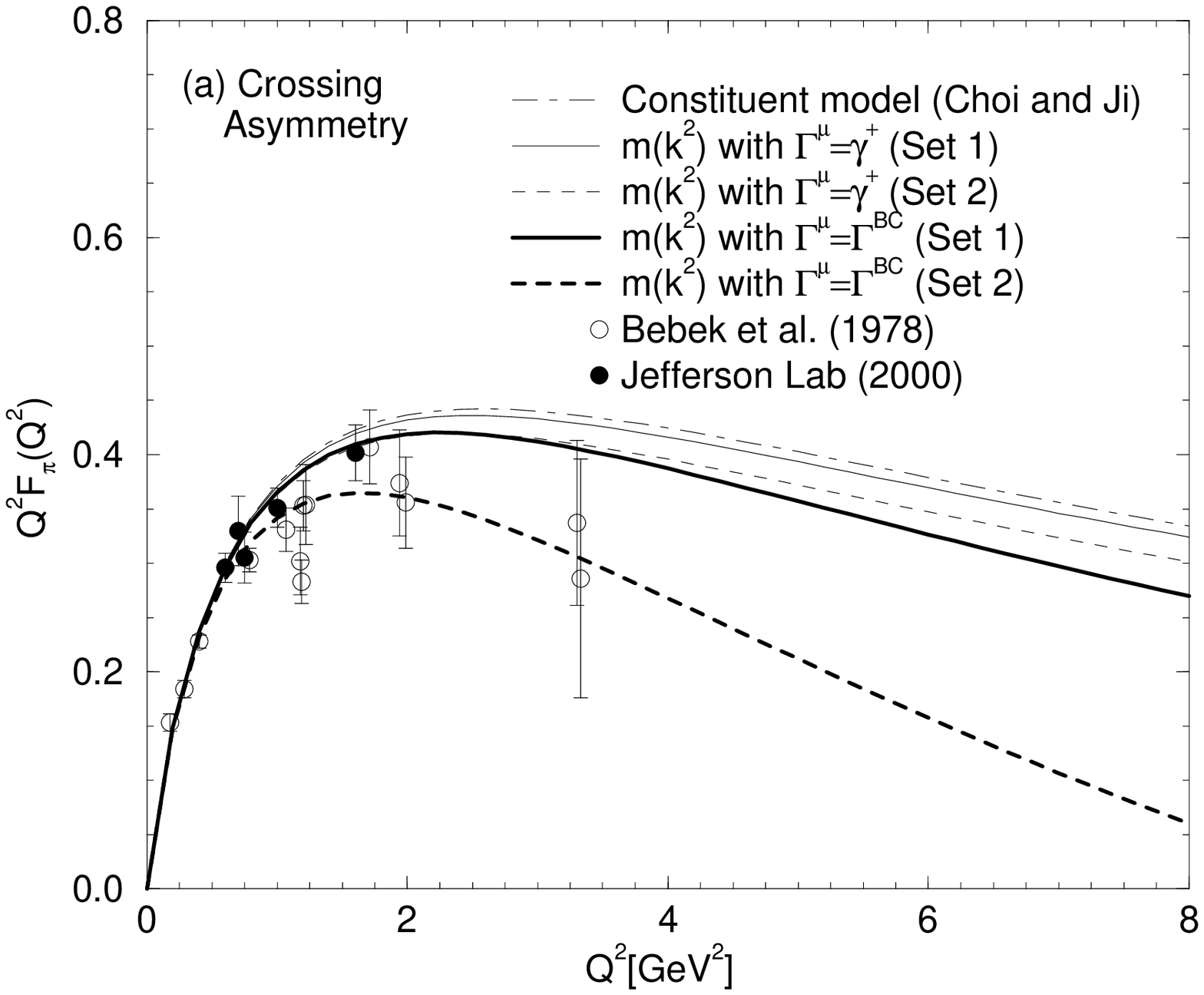,height=70mm,width=70mm}}
\vspace{-1.2cm}
\centerline{\epsfig{figure=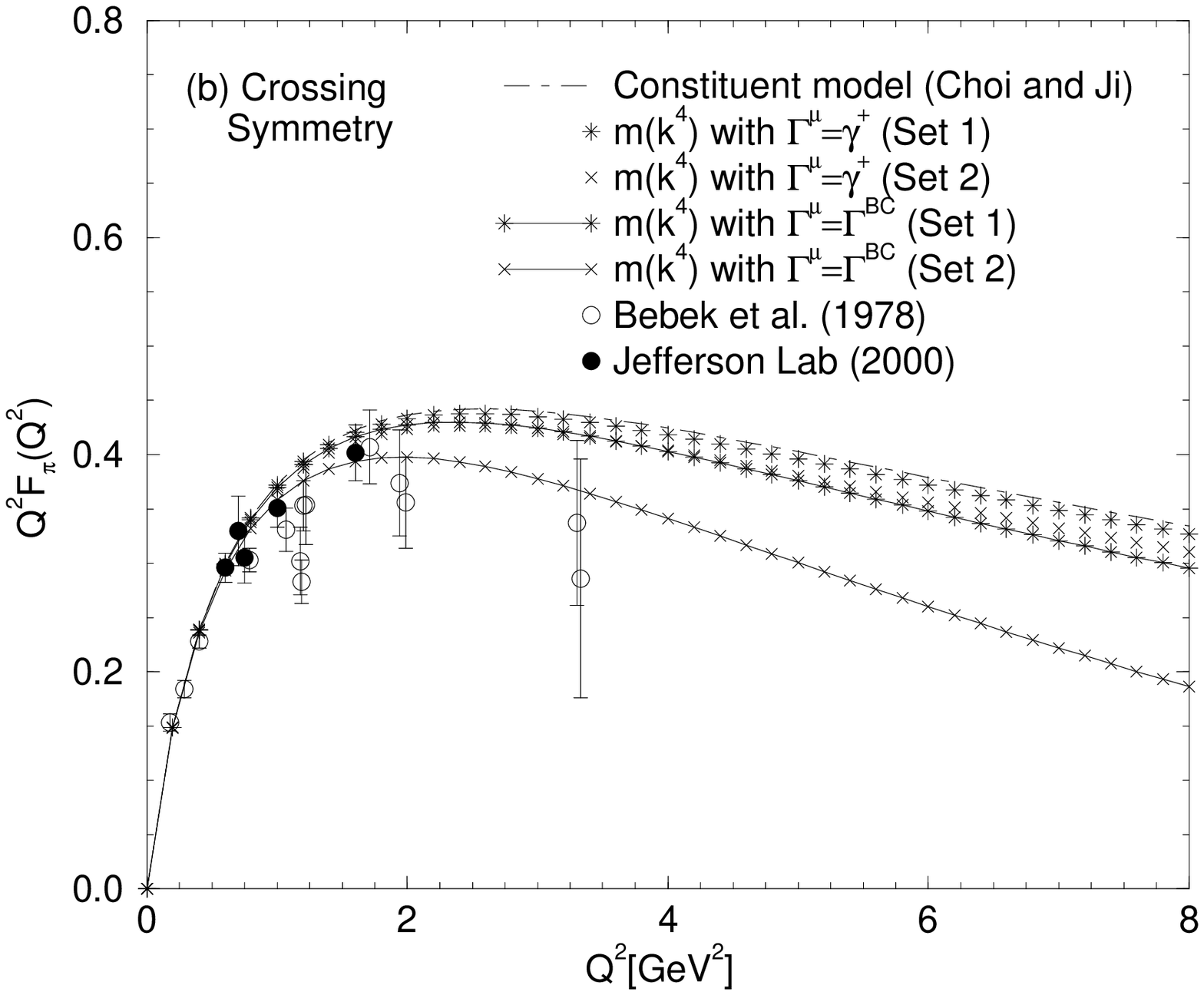,height=70mm,width=70mm}}
\vspace{-1.2cm}
\caption{Pion EM form factor:(a) Crossing asymmetry(CA) and
(b) Crossing symmetry(CS) mass functions compared with the experimental
data~\protect\cite{Bebek} as well as the CQM result~\protect\cite{CJ}.
\label{pi1}}
\end{figure}
In Fig.~\ref{pi1}, we show our results of the form factor
for the intermediate $Q^2$ region for CA [Fig.~\ref{pi1}(a)]
and CS  [Fig.~\ref{pi1}(b)] mass functions compared with
the experimental data~\cite{Bebek} as well as the CQM result~\cite{CJ}.
As one can see from Fig.~\ref{pi1}, 
(1) there are differences between the
bare vertex and BC ansatz indicating the breakdown of local gauge
invariance from the usage of the bare vertex,
(2) the [Set 2] for both CA and CS mass functions show larger deviation
from the CQM result than the [Set 1] case for 
the region of momentum transfer
$Q^2\sim$2 GeV$^2$ and above,
(3) the results with the BC vertex fall off faster (at around $Q^2$=2 GeV$^2$)
than the CQM result does,
and (4) the CA mass evolution function is more
sensitive to the variation of the momentum dependence than the CS mass
evoluton function.

In conclusion, we have reexamined the soft contribution to the
pion elastic form factor using LFQM with a running quark mass.
The Ball-Chiu ansatz was used for the dressed  quark-photon vertex.
We were also able to calculate the quark condensate as
$-\la{\bar q}q\ra$=(0.3 GeV)$^3$~\cite{KCJ}, which is quite compariable with
the value employed in contemporary phenomenological studies:
(0.236 GeV)$^3$~\cite{qqE}.
This shows the PCAC relation is reasonably well satisfied in LFQM
with our mass evolution function.

\end{document}